\documentclass{PoS}

\title{HAWC response to atmospheric electricity activity}

\ShortTitle{Atmospheric Electricity and HAWC}

\author{\speaker{Alejandro Lara}\\
        Instituto de Geof\'isica, Universidad Nacional Aut\'onoma de M\'exico\\
        E-mail: \email{alara@igeofisica.unam.mx}}
\author{Graciela Binimelis de Raga \\
        Centro de Ciencias de la Atm\'osfera, Universidad Nacional Aut\'onoma de M\'exico\\}
\author{Olivia Enr\'iquez-Rivera\\
        Instituto de Geof\'isica, Universidad Nacional Aut\'onoma de M\'exico\\
        }

\author{for the HAWC collaboration}
\abstract{The HAWC Gamma Ray observatory consists of 300 water Cherenkov detectors (WCD) instrumented with four photo multipliers tubes (PMT) per WCD. HAWC is located between two of the highest mountains in Mexico. The high altitude (4100 m asl), the relatively short distance to the Gulf of Mexico (\textasciitilde100 km), the large detecting area (22 000 m$^2$) and its high sensitivity, make HAWC a good instrument to explore the acceleration of particles due to the electric fields existing inside storm clouds. In particular, the scaler system of HAWC records the output of each one of the 1200 PMTs as well as the 2, 3, and 4-fold multiplicities (logic AND in a time window of 30 ns) of each WCD with a sampling rate of 40 Hz. Using the scaler data, we have identified 20 enhancements of the observed rate during periods when storm clouds were over HAWC but without cloud-earth discharges. These enhancements can be produced by electrons with energy of tens of MeV, accelerated by the electric fields of tens of kV/m measured at the site during the storm periods. In this work, we present the recorded data, the method of analysis and our preliminary conclusions on the electron acceleration by the electric fields inside the clouds.}

\FullConference{35th International Cosmic Ray Conference  ICRC217-\\
		10-20 July, 2017\\
		Bexco, Busan, Korea}

\usepackage{lineno}
\begin{document}
%\linenumbers

\section{Introduction}

Particle acceleration up to high energies inside the Earth's atmosphere
has been observed by satellite gamma ray detectors
during terrestrial gamma ray flashes \cite{Fishman,Smith,Marisaldi,Briggs}.
At ground level, high altitude cosmic ray detectors have reported
ground enhancements during thunderstorms
\cite{Alexeenko,Vernetto,Chilingarian,Tsuchiya}.

The development of large electric fields during thunderstorms (up to
200 kV/m \cite{Stolzenburg}) accelerates charged
particles. Electrons may gain energies up to tens of MeV \cite{Chilingarian}. In this work we report enhancements of the count rates observed by
the High Altitude Water Cherenkov Observatory (HAWC) which might be related to the
atmospheric electric field.

HAWC is an air shower detector located at 4,100 m a.s.l, N $18^{\circ} 59' 48''$, W  $97^{\circ} 18' 34''$.  Built on the slope of Sierra Negra, Puebla in Mexico it consists of 300 water Cherenkov detectors 7.3 m diameter and 4.5 m deep. Each tank is filled with filtered water and the total detector comprises an extension of $22,000$ m$^2$.

HAWC is operating on one of the highest mountains in
Mexico. HAWC's high altitude together with the proximity of the Gulf of
Mexico make the array an excellent
laboratory to study the high energetic processes during thunderstorms. In section \ref{sec:site} we discuss the weather on the HAWC site in more detail.

In section \ref{sec:scaler} we describe HAWC  scaler systems which are
able to detect low energy particles. 
In Sec. \ref{sec:enhan} we
present the enhancements of the scaler count rates  observed by HAWC due
to the presence of strong electric fields. Finally our discussion is presented in Sec. \ref{sec:disc}

\section{The weather at the HAWC site}  \label{sec:site}

Southern and Central Mexico is located in the tropics, with ample humidity during most of the year and characterized by a 6-month rainy season. HAWC in particular, is located in a region with frequent presence of clouds formed by forced orographic lifting or due to atmospheric convective instability.  The latter mechanism often leads to cumulus clouds responsible for the development of precipitation and are also responsible for charge separation within the cloud, due to collisions between hydrometeors at the different temperature ranges observed.   The development of poles of positive and negative charge within the cloud give rise to an electric field that can reach breakdown point and result in a lightning discharge, within the cloud or from the cloud to ground.

Several ground-based networks have been developed to monitor cloud-to-ground lightning continuously to assess the risk to the population. One such global network is the World Wide Lightning Location Network (WWLN), documented by Dowden et al (2008). The first studies of combined precipitation and lightning over Mexico were carried out in 2010 and revealed the regions of the country where most of the cloud-to-ground lightning is observed (Kucienska et al, 2010). Furthermore, Raga et al (2014) showed that Mexico is particularly vulnerable, with a large number of deaths per year.

While HAWC is not located in the region of highest incidence of lightning in Mexico, its location on the Sierra Negra ensures that it will be affected by electrically-charged clouds and lightning for about 6 months of the year.  The electric fields associated with the convective clouds and the proximity with intra-cloud and cloud-to-ground lightning provide a unique setting to study the effect of these phenomena on the measurements made by HAWC. In this work, we carried out measurements of the ambient variables using a weather station placed at the HAWC site. To measure the electric field, we used an electric field mill Boltek-100 installed at the eastern side of the array. Plots of these variables are shown in section \ref{sec:enhan}.

\begin{figure}[h]
\centerline{\includegraphics[width=0.8\textwidth]{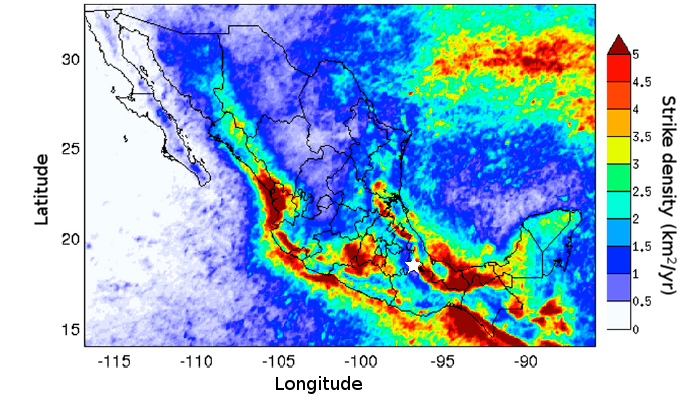}}
\caption{Average spatial distribution of cloud-to-ground lightning density in flashes per square kilometer per year for the period 2006-2012, adapted from Raga et al (2014). The white star mark the location of HAWC.} \label{fig:mapa}
\end{figure}

\section{HAWC scaler systems} \label{sec:scaler}

HAWC data are collected by two data acquisition systems (DAQs). The main DAQ measures arrival times and time over thresholds of PMT pulses and allows for the reconstruction of the air shower arrival direction and energy of the primary particle. The electronics are based on time to digital converters (TDC). The main DAQ also has a TDC scaler system which counts the hits inside a time window of 30 ns of each PMT and the coincidences of 2, 3 and 4 PMTs in each water Cherenkov detector. These coincidences are called multiplicity 2, 3 and 4, respectively. The secondary DAQ consists of a counting system that registers each time the PMT is hit by $>1/4$ photoelectron charge and we call it hardware (HW) scaler system. This simpler system together with the TDC scalers allows one to measure particles below the energies of reconstructable showers.

\section{Count rate enhancements} \label{sec:enhan}

\begin{figure}[h]
\centerline{\includegraphics[width=0.85\textwidth]{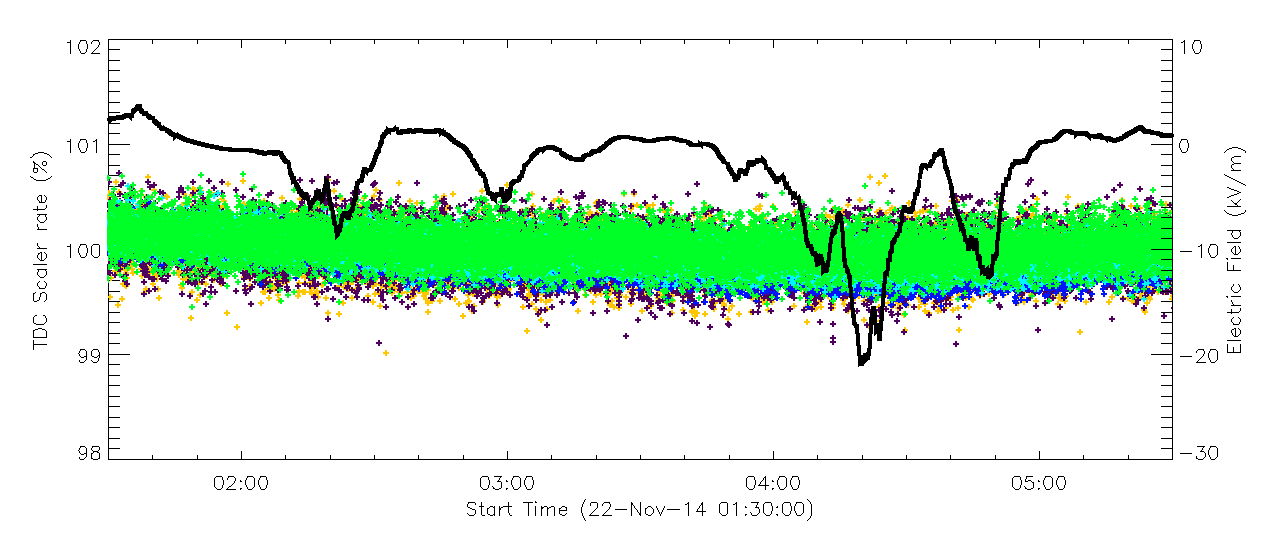}}
\caption{Mean count rate of the TDC scaler Multiplicities: 2 (blue),
  3 (cyan) and 4 (green); and rates of the 8'' (orange) and 10'' (purple) PMTs, during
  November 22, 2014 when a moderate negative electric field was
  observed (black curve).} \label{fig:quiet}
\end{figure}

\begin{figure}[h]
\centerline{\includegraphics[width=0.85\textwidth]{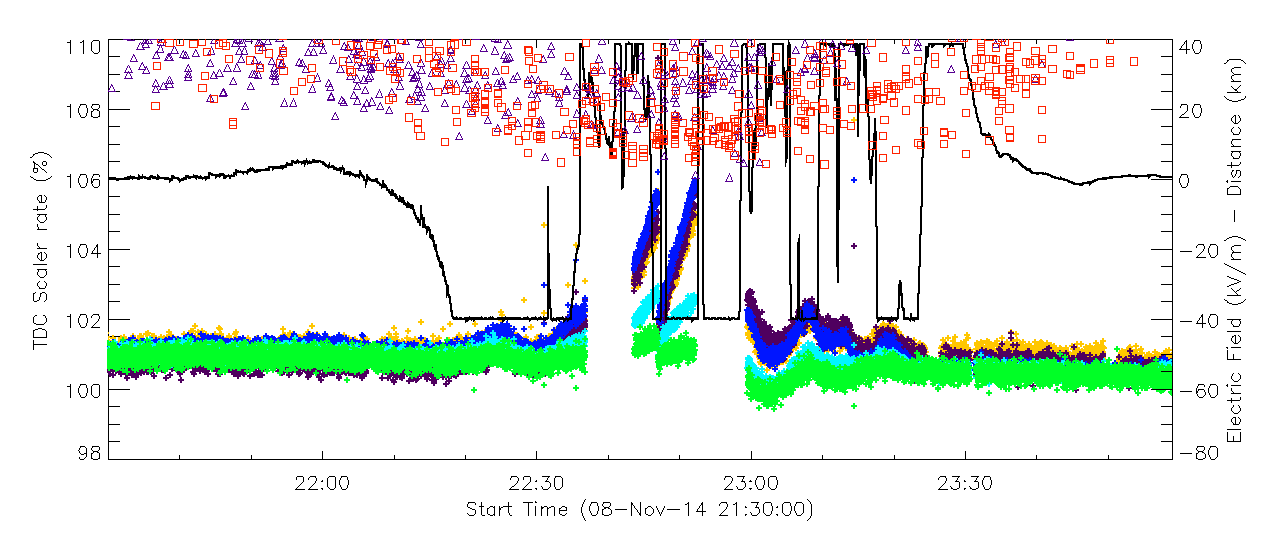}}
\caption{TDC scaler rates and electric field measured during Nov 8, 2014.
  The color code is similar as Fig. \ref{fig:quiet}. The distance of the reported cloud to ground (red squares) and inter-cloud (purple triangles) discharges are also plotted.} \label{fig:disturb} 
 \end{figure}

We have noted that 
the HAWC scaler systems  responds to the atmospheric electricity
at least in four  ways: i) when the electric field is positive or weak
negative the count rate does not suffer any change as seen in Figure
\ref{fig:quiet} where we have plotted the TDC scaler rates during the negative
electric field  enhancement observed on Nov 22, 2014. All scaler rates are in
percentage taking as reference, i.e. 100\% , the mean scaler rate calculated one hour before the event started. The electric field is shown in black solid line. As an eye aid, the equivalent zero electric field
is plotted as a horizontal dashed line.
ii) when the storm is very close to the array with large amount
of discharges, the system behavior is unstable and the system restarts
frequently. 
There are scaler enhancements but they may be due to the discharges and/or
the electric field or electromagnetic noise.
For example, Figure \ref{fig:disturb} shows
the TDC scaler rates during November 8, 2014, an active day in terms of
atmospheric electricity. A thunderstorm took place during this period as seen by the red square symbols and purple triangles representing the cloud to ground and inter-cloud discharges, respectively. There were nearby discharges as shown by the rapid variations of the field strength (it is important to note that the
electric field detector gets saturated when the electric field is larger
than $\pm 40$ kV/m).

\begin{figure}[h]
\centerline{\includegraphics[width=0.85\textwidth]{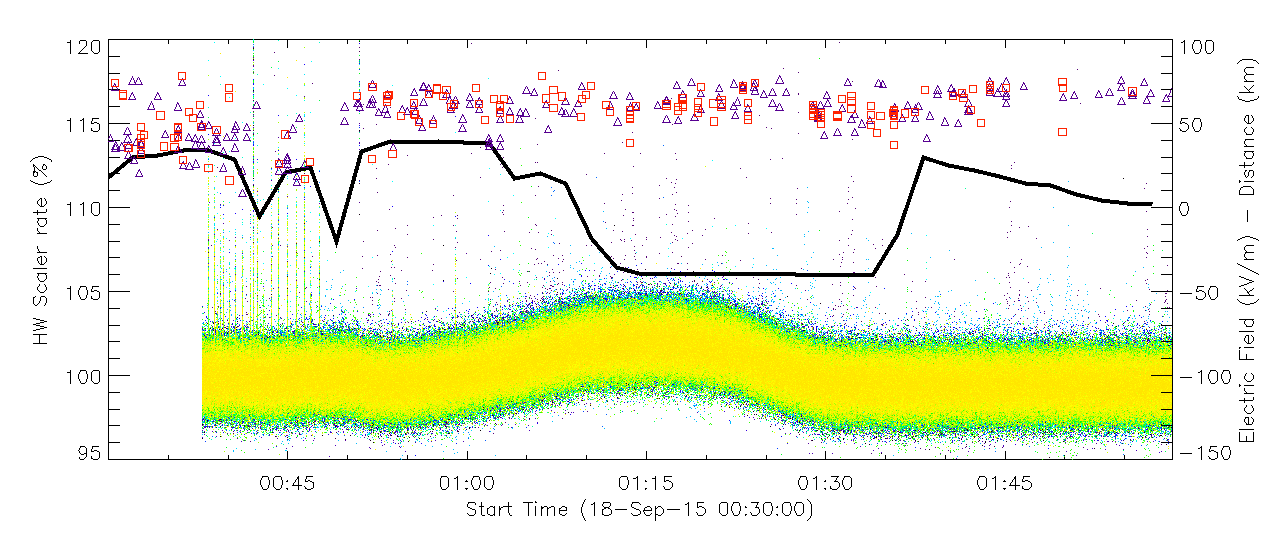}}
\caption{Mean count rate of the HW scaler system,  each available
  channel is plotted in colored dots. The distance of the reported cloud
 to ground (red squares) and inter-cloud (purple triangles) discharges
 are also plotted.} \label{fig:medium}
\end{figure}

There are some events where the  atmospheric electric activity is not so strong
and therefore, the response of the scaler system is somehow ``well behaved.''
In those cases we can distinguish:
iii) a fast response of the scaler system associated  
to the discharges and
iv) a slow response to strong negative electric fields.
These responses are depicted in Figure \ref{fig:medium} where we have plotted, 
with colored dots, the count rate (in percentage) of each channel of the
HW scaler system during
September 18, 2015. The electric field shows two rapid changes around 00:45,
associated with discharges. The squares and triangles indicate that a storm took place. There are sharp scaler enhancements associated with the closest discharges
between 00:35 and 00:50 UT.

\begin{figure}[h]
\centerline{\includegraphics[width=0.85\textwidth]{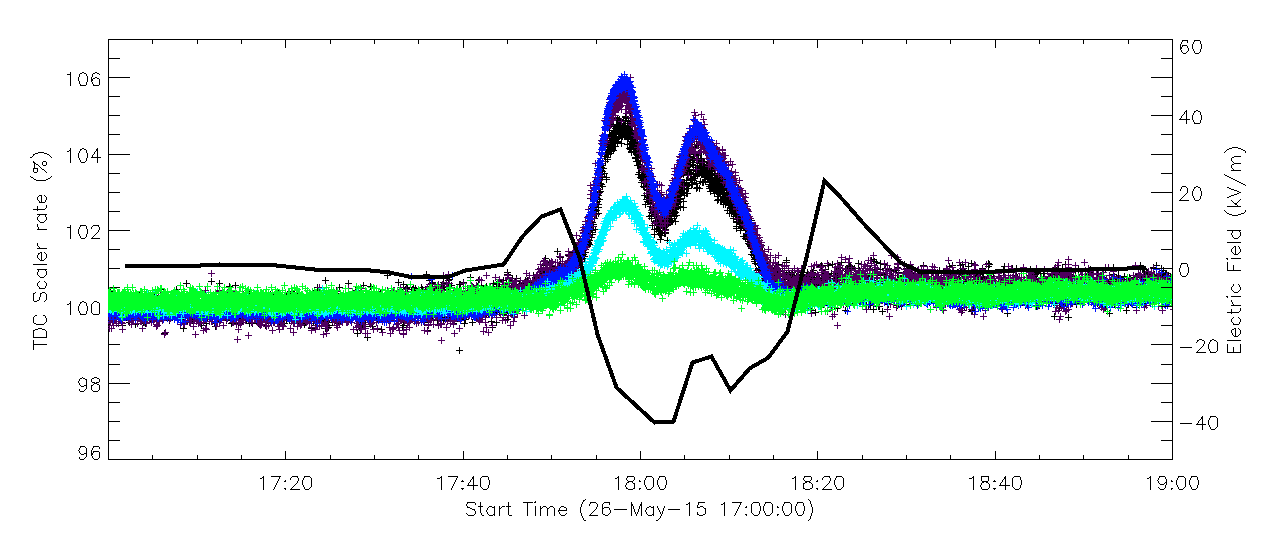}}
\caption{Mean count rate of the TDC scaler Multiplicities during May 26,
  2015. The color code is the same as in Figure \ref{fig:quiet}.} \label{fig:tdcmay26}
\end{figure}

\begin{figure}[h]
\centerline{\includegraphics[width=0.85\textwidth]{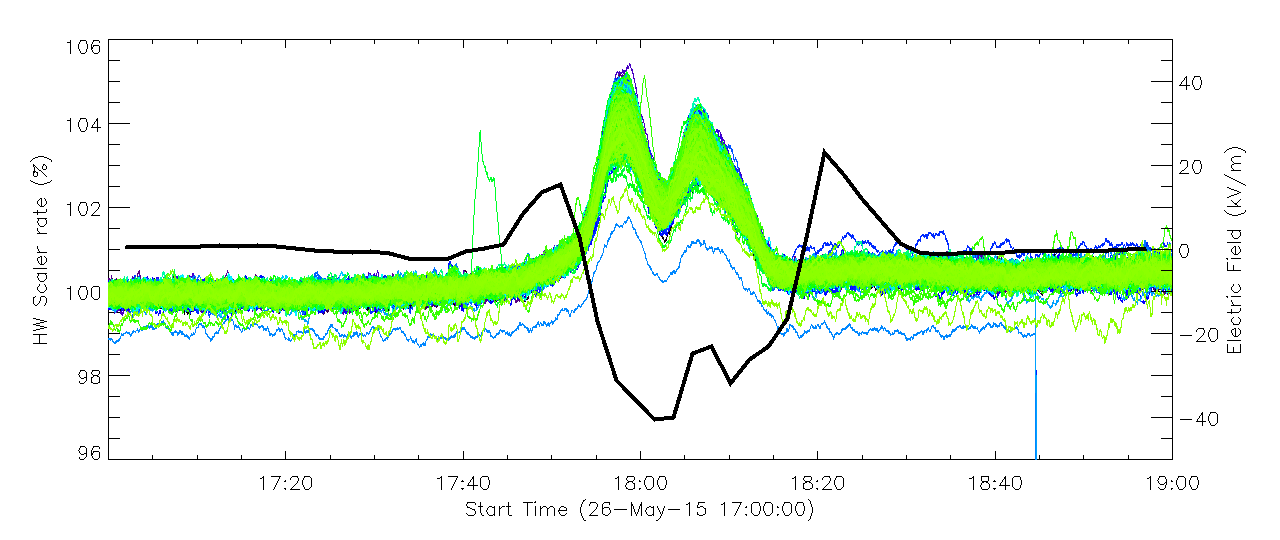}}
\caption{Mean count rate of the HW scaler system during May 26, 2015. The color code is the same as in Figure \ref{fig:medium}.} \label{fig:hwmay26}
\end{figure}

In this work we focus on the slow response of the scaler system due to the
presence of strong negative electric field.
A clear example of this slow response was observed during May 26, 2015
and is depicted in Figures \ref{fig:tdcmay26} and \ref{fig:hwmay26} for TDC and HW scalers,
respectively. We have selected  this event due to the fact that there is no
saturation of the electric field measurements.
%We have selected this event because there is no saturation
%of the electric field measurements.
%

\begin{figure}[h]
\centerline{\includegraphics[width=0.85\textwidth]{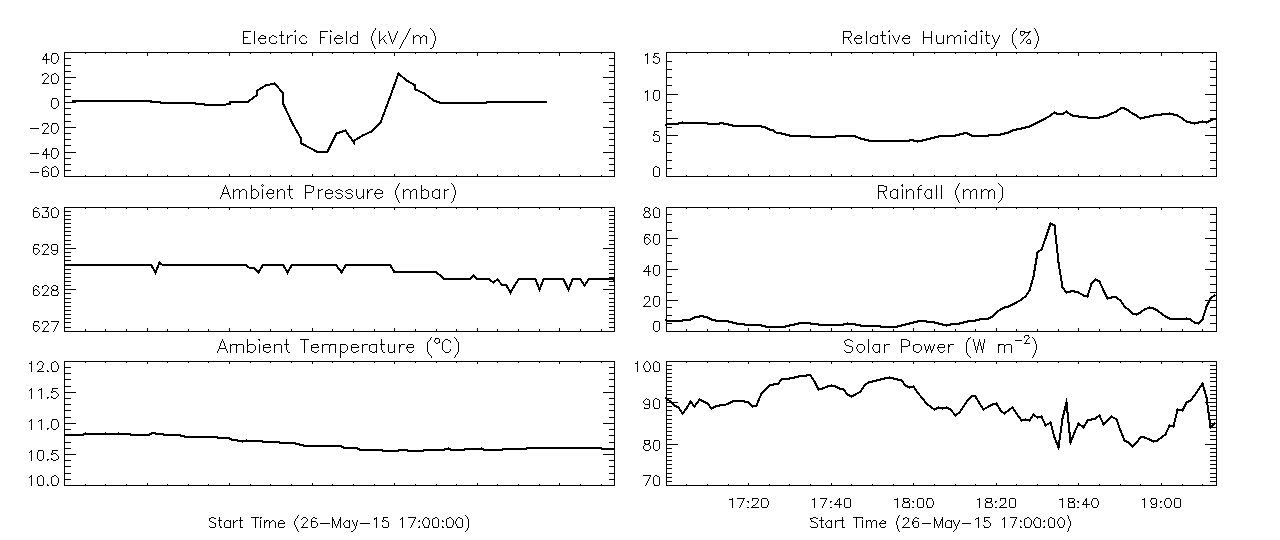}}
\caption{Environment variables at the HAWC site during May 26, 2015. From top to bottom are:
  Electric field, ambient pressure, temperature (left panels),
  humidity, rainfall and solar irradiance (right panels).} \label{fig:envmay26}
\end{figure}

The environment parameters during May 26, 2015 are presented in
Figure \ref{fig:envmay26}. From top to bottom, we plotted the electric field, pressure, temperature, humidity, rain fall and solar irradiance. The latter two are displayed in order to show the presence of clouds at the site during the scaler enhancements.  In particular, it is
well known that the
count rate of cosmic ray detectors has an inverse dependence on the
ambient pressure. This figure shows that the count rate
enhancements during the events are not related to pressure changes.

\begin{figure}[h]
\centerline{\includegraphics[width=0.85\textwidth]{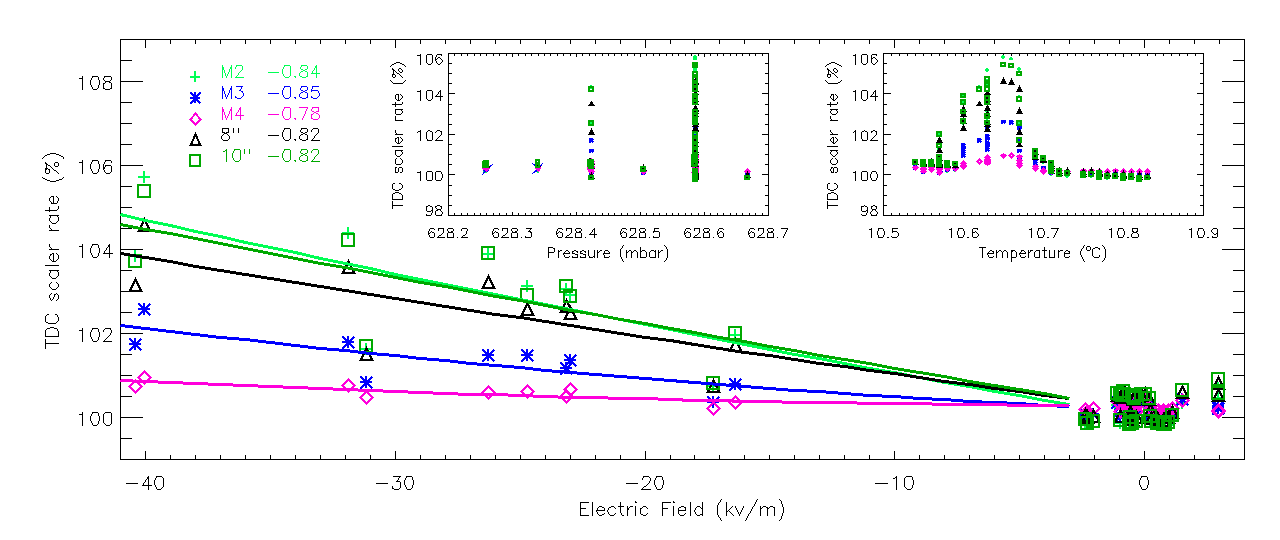}}
\caption{Scatter plot of the TDC scale Multiplicity 2 (green), 3 (blue)
  4 (magenta); and 8'' (black) and 10'' PMT rates as a function of the
  electric field during May 26, 2015. The correlation coefficients are shown next to each Multiplicity and PMT set. Upper panels: scatter plots of the scaler rates vs Pressure (left) and Temperature (right)} \label{fig:corrmay26}
\end{figure}

Figure \ref{fig:corrmay26} shows the scatter plot of the TDC scaler
enhancement as a function of the strength of the electric field,
during the May 26 event. Unfortunately the electric field data are stored with poor time resolution ($\sim$ 1.5 minutes), limiting the statistics available for correlations. In order to show the tendency of the correlation, 
we fit a second degree polynomial to each mean multiplicity and 
mean PMT rates. The correlation coefficients are shown in the plot as reference. The scatter is high but one can see that both 10'' PMT and
multiplicity 2 rates are more affected by the electric field enhancement.
The 8'' PMTs, multiplicity 3 and finally multiplicity 4 are less affected.
If the rate enhancements are being produced by the acceleration of charged particles in the electric field, this correlation would indicate more low-energy and few high-energy particles in the enhancement. Finally, the small scatter plots in Figure \ref{fig:corrmay26} emphasize the lack of correlation between the scaler rates and the Pressure/Temperature measured at the site.

\section{Discussion} \label{sec:disc}
In this work, we presented examples of the HAWC scaler system response to the atmospheric
electricity activity. 
We showed an example of  the rapid response of the HW scaler system
 due to a close lightning
activity (Figure \ref{fig:medium}). However, in this work
we focus in the scaler system response  to the negative electric
field. 

In particular we presented an example in which the electric
field was not saturated and was observed by the two scaler systems.
The correlated enhancement of all the available PMTs of the array seen
by the HW scaler system (Figs. \ref{fig:hwmay26})
shows that the enhancement embraces the entire array with similar
response at time scales of seconds. The preliminary correlation analysis between the negative electric
field and TDC count rates shows the high relationship between
these variables (Figures \ref{fig:corrmay26}). Furthermore,  the absence of correlation between the scaler count rates and  atmospheric variables such as pressure or temperature supports a possible scenario where the scaler rate enhancements might be produced by particle acceleration due to the electric field of clouds observed by HAWC. If our hypothesis is correct, the enhancements of all multiplicities of the TDC scaler system (Fig. \ref{fig:tdcmay26}) will allow us to determine the energy of the incident particles.

We have shown that HAWC can be a  good instrument to study the
acceleration of particles by  the atmospheric electricity. It is necessary to perform a detailed analysis and simulations to quantify our observations, as well as rule out instrumental effects in the photomultipliers (such as inductive charging) that could be producing the observed rate enhancements. This analysis will be published elsewhere.

{\bf Acknowledgments}

We acknowledge the support from: the US National Science Foundation (NSF); the US Department of Energy Office of High-Energy Physics; the Laboratory Directed Research and Development (LDRD) program of Los Alamos National Laboratory; Consejo Nacional de Ciencia y Tecnolog\'ia (CONACyT), M\'exico (grants 271051, 232656, 260378, 179588, 239762, 254964, 271737, 258865, 243290, 132197), Laboratorio Nacional HAWC de rayos gamma; L'OREAL Fellowship for Women in Science 2014; Red HAWC, M\'exico; DGAPA-UNAM (grants IG100317, IN111315, IN111716-3, IA102715, 109916, IA102917); VIEP-BUAP; PIFI 2012, 2013, PROFICIE 2014, 2015; the University of Wisconsin Alumni Research Foundation; the Institute of Geophysics, Planetary Physics, and Signatures at Los Alamos National Laboratory; Polish Science Centre grant DEC-2014/13/B/ST9/945; Coordinaci\'on de la Investigación Cient\'ifica de la Universidad Michoacana. Thanks to Luciano D\'iaz and Eduardo Murrieta for technical support.

\end{document}